\documentclass[lettersize,journal]{IEEEtran}
\usepackage{amsmath,amsfonts}
\usepackage{algorithmic}
\usepackage{algorithm}
\usepackage{array}
\usepackage[caption=false,font=normalsize,labelfont=sf,textfont=sf]{subfig}
\usepackage{textcomp}
\usepackage{stfloats}
\usepackage{url}
\usepackage{verbatim}
\usepackage{enumitem}
\usepackage{graphicx}
\usepackage{cite}
\usepackage{xspace}
\DeclareMathOperator*{\argmin}{arg\,min}
\usepackage{tabularray}
\UseTblrLibrary{booktabs}
\SetTblrInner{rowsep=1pt}

\usepackage[resetlabels]{multibib}
\usepackage{chngcntr}
\newcites{SM}{References}

\usepackage[pagebackref=true,breaklinks=true,letterpaper=true,colorlinks,bookmarks=false,citecolor=blue]{hyperref}
\usepackage{cleveref}

\newcommand{\opencam}{\textsc{OpEnCam}\xspace}

\begin{document}

% \title{\opencam: Designed Lensless Cameras for Optical Encryption}
\title{\opencam: Lensless Optical Encryption Camera}

\author{Salman S. Khan, Xiang Yu, Kaushik Mitra, Manmohan Chandraker, Francesco Pittaluga
        % <-this % stops a space
\thanks{S. S. Khan is with Rice University, Houston, TX, USA - 77005}
\thanks{X. Yu is with Amazon, USA}
\thanks{K. Mitra is with the Indian Institute of Technology Madras, Chennai, India - 600036 }
\thanks{M. Chandraker and F. Pittaluga are with the NEC Labs America, San Jose, CA, USA - 95110}
\thanks{This paper was produced by the IEEE Publication Technology Group. They are in Piscataway, NJ.}}% <-this % stops a space}



\maketitle

\begin{abstract}
Lensless cameras multiplex the incoming light before it is recorded by the sensor. This ability to multiplex the incoming light has led to the development of ultra-thin, high-speed, and single-shot 3D imagers. Recently, there have been various attempts at demonstrating another useful aspect of lensless cameras - their ability to preserve the privacy of a scene by capturing encrypted measurements. However, existing lensless camera designs suffer numerous inherent privacy vulnerabilities. To demonstrate this, we develop the first comprehensive attack model for encryption cameras, and propose \opencam -- a novel lensless \text{op}tical \textbf{en}cryption \textbf{ca}mera design that overcomes these vulnerabilities. \opencam encrypts the incoming light before capturing it using the modulating ability of optical masks. Recovery of the original scene from an \opencam measurement is possible only if one has access to the camera's encryption key, defined by the unique optical elements of each camera. Our \opencam design introduces two major improvements over existing lensless camera designs - (a) the use of two co-axially located optical masks, one stuck to the sensor and the other a few millimeters above the sensor and (b) the design of mask patterns, which are derived heuristically from signal processing ideas. We show, through experiments, that \opencam is robust against a range of attack types  while still maintaining the imaging capabilities of existing lensless cameras. We validate the efficacy of \opencam using simulated and real data. Finally, we built and tested a prototype in the lab for proof-of-concept.
\end{abstract}


%%%%%%%%%%%%%%%%%%%%%%%%%%%%%%%%%%%%%%%%%%%%%%%%%%%%%%%%%%%%%%%%%%%%%%%%%%
\section{Introduction}
%%%%%%%%%%%%%%%%%%%%%%%%%%%%%%%%%%%%%%%%%%%%%%%%%%%%%%%%%%%%%%%%%%%%%%%%%%

\begin{figure}[t]
    \centering
    \includegraphics[width=\linewidth]{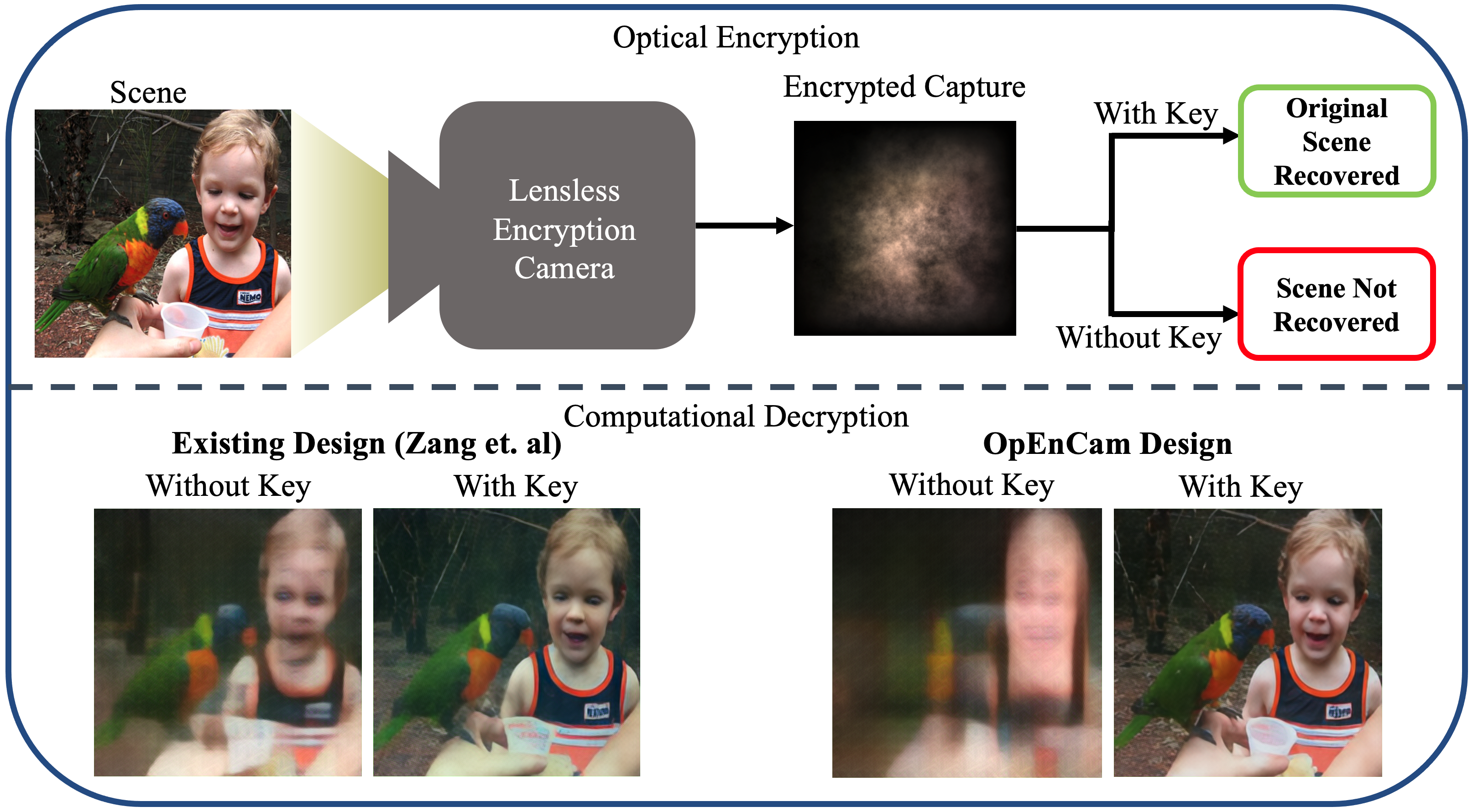}
   \caption{\textbf{Optical Encryption.} Optical encryption camera encodes the scene optically prior to capture. Decryption is possible computationally using a key that is unique to the hardware of each encryption camera. Existing optical encryption designs are susceptible to powerful learning-based attacks even without the key. Proposed \opencam design prohibits such learning-based attacks effectively.}
    \label{fig:teaser}
\end{figure}

Lensless imaging is a computational imaging modality that replaces the lens of a conventional camera with a thin optical mask and computation. The addition of the mask, which no longer has the focusing ability of the lens, leads to multiplexing of the incoming light prior to capture. This multiplexing of light has been exploited to develop ultra-thin cameras for 2D \cite{asif2016flatcam,boominathan2020phlatcam} and 3D imaging \cite{boominathan2020phlatcam,antipa2018diffusercam} high-speed imaging \cite{antipa2019video} and hyper-spectral imaging \cite{monakhova2020spectral}. Another interesting aspect of lensless cameras that has been relatively less explored is their ability to preserve privacy in the optical domain itself before capture. This is made possible by the ability of lensless cameras to perform computation in the optical space, with operations defined by their unique mask patterns and arrangements. 

With camera-based technologies becoming increasingly integrated into every aspect of our lives, the potential for leakage of sensitive visual information is ever-increasing. This is leading to significant privacy and security concerns from  consumers, citizens, and governments. The standard approach to address these concerns is to encrypt sensitive image data at the sensor level, after image capture, via software or specialized hardware. However, most cameras lack such specialized hardware. Further, both software and hardware-based solutions may still be susceptible to data sniffing attacks that gain access to sensitive data before encryption. Thus, protecting the privacy of a scene through optical processing before capture is currently the need of the hour.

While optical image encryption has been around for many decades \cite{refregier1995optical}, most existing solutions require coherent illumination for encoding and decoding, making them unsuitable for deployment in unconstrained real-world environments. Further, incoherent illumination solutions that have been proposed are susceptible to ciphertext-only attacks\cite{Zang:13} or have yet to be realized with real hardware due to the experimental nature of the optics\cite{Byrne2020bmvc}.

A few works like \cite{nguyen2019deep,ishii2020privacy,wang2019privacy} have used existing lensless cameras \cite{asif2016flatcam} to show optical encryption-based privacy-preserving applications. However, these works are fundamentally limited by the inherent design flaws of the existing lensless cameras, which were designed for imaging applications with no privacy considerations. In this work, we look at the design of existing lensless cameras and make two significant modifications to it that allows us to improve their privacy-preserving ability prior to capture. Our novel lensless camera design is called \opencam – Optical Encryption Camera. In \opencam, we first introduce a novel double mask design using two co-axially placed masks – one at the top of the bare sensor placed flush against it and the other a few millimeters above it. Such a double mask design makes lensless cameras significantly less susceptible to various attacks without compromising their imaging ability. Secondly, we introduce a novel design strategy based on signal processing heuristics for the two masks used in our system that further enhances the privacy of our system. Finally, we show through experiments that our proposed \opencam design still has the imaging ability of existing lensless cameras while preserving privacy.

To summarize, we make the following contributions:
\begin{enumerate}%[noitemsep,nolistsep]
\item A novel framework for generating lensless computational cameras that encrypt image data before image capture, using two co-axially placed masks - a scaling mask placed flush against the bare sensor, and a multiplexing mask placed a few millimeters from the sensor. 
\item Novel heuristically designed random key-generators for the two masks, providing a distinct mask pattern for each camera. We show that these key-generators are robust against various forms of ciphertext-only attacks and known plaintext attacks.
\item A novel transformer-based keyed decryption approach.
\item We validate our results both in simulation and real data collected using an inexpensive proof-of-concept prototype built in our lab.
\end{enumerate}

\section{Background}
\subsection{Optical Cryptography}
%%%%%%%%%%%%%%%%%%%%%%%%%%%%%%%%%%%%%%%%%%%%%%%%%%%%%%%%%%%%%%%%%%%%%%%%%%

A cryptographic system is judged to be reliably secure only if it can successfully survive a rigorous evaluation via cryptanalysis. The goal of cryptanalysis is to identify potential weaknesses in encryption mechanisms that would allow someone to decode the encrypted data, even without possessing the confidential key. Technological progress in the realm of optical cryptography has spurred the introduction of various specialized cryptanalysis approaches, such as chosen plaintext attacks (CPA) \cite{peng2006chosen,liao2017optical}, known plaintext attacks (KPA) \cite{peng2006known,gopinathan2006known}, and ciphertext-only attacks (COA) \cite{peng2007ciphertext,zhang2013ciphertext,liu2015vulnerability}. 
% At the same time, understanding the vulnerabilities in existing optical encryption methods can contribute to the design of more fortified systems \cite{sahoo2017enhancing,liao2017security,cheng2008security}.

In a CPA scenario, the attacker can choose arbitrary plaintexts to be encrypted and then examine the resulting ciphertexts. The primary objective is to use this information to discover a weakness in the encryption algorithm or even to deduce the secret key used for encryption. In contrast, in a KPA scenario, the attacker does not have the luxury of choosing the plaintexts that are encrypted. Instead, the attacker must work with pre-existing pairs of plaintexts and ciphertexts. Despite this limitation, a successful KPA could potentially reveal enough about the encryption scheme or key to decrypt other ciphertexts encrypted with the same key or to weaken the overall security of the cryptosystem. Finally, in a COA, the attacker has access solely to the ciphertext—that is, the encrypted data—without any accompanying plaintext or other additional information. The goal of the attacker is to deduce either the plaintext, the encryption key, or details about the encryption algorithm, based solely on the available ciphertexts. Ciphertext-only attacks are considered one of the most challenging forms of cryptanalysis because the attacker has minimal information to work with. Unlike chosen CPA or KPA, where the attacker has more control or information, COA provides the least amount of leverage for the attacker. If a cryptographic algorithm can resist a ciphertext-only attack, it's generally considered to be quite secure.

One of the earliest works on optical encryption is \cite{refregier1995optical}. The authors place independent white uniformly distributed phase masks in the input and Fourier planes of a 4f correlator to encode images.  However, this approach only works for coherent illumination, and is also susceptible to autocorrelation-based COAs and impulse-based CPAs \cite{liao2021deep}. \cite{liao2021deep} improves on \cite{refregier1995optical} by adding a third independent white uniformly distributed phase mask in the image plane of the 4f correlator to inhibit impulse-based CPAs \cite{liao2021deep}. In \cite{liao2017ciphertext}, the authors place a random white uniformly distributed phase mask in the aperture plane of a camera to produce an incoherent imaging system for optical image encryption and demonstrate that such a system is susceptible to autocorrelation-based COAs. In \cite{Byrne2020bmvc}, the authors simulate an experimental 3D printed optical fiber bundle \cite{wang20183d} at the imaging plane of a camera to produce pixel shuffling. However, such optical fiber bundles have only been shown for parallel fibers, which are not suitable for pixel shuffling \cite{wang20183d}. Our framework employs standard optical masks that are cost-effective and easily mass-produced, and we construct a prototype encryption camera. Works like \cite{ishii2020privacy,Zang:13,wang2019privacy,nguyen2019deep} use single optical masks for encryption. Although these methods are practical, they are not robust to various decryption attacks. In comparison, our solution is not only practical but also robust to these attacks. It should also be noted that none of the above existing works provide a complete analysis of the different forms of attacks that need to be considered for an optical encryption system, especially the powerful learning-based attacks that exploit hidden priors in data. 

%%%%%%%%%%%%%%%%%%%%%%%%%%%%%%%%%%%%%%%%%%%%%%%%%%%%%%%%%%%%%%%%%%%%%%%%%%
\subsection{Lensless Imaging}
%%%%%%%%%%%%%%%%%%%%%%%%%%%%%%%%%%%%%%%%%%%%%%%%%%%%%%%%%%%%%%%%%%%%%%%%%%

In a conventional single-mask lensless camera like \cite{asif2016flatcam,boominathan2020phlatcam}, the lens of the camera is replaced by a thin optical mask placed at some distance from the sensor that modulates the incoming light. For a scene $X$, the measurement recorded by a sufficiently large sensor $Y$ is given by:
\begin{equation}
    Y = P * X + N,
\end{equation}
where $*$ is the full-size convolutional operator (no cropping due to finite sensor size),  $P$ is the point spread function (PSF), and $N$ is additive noise. The PSF is the response of the camera to a point source, and it depends on the mask pattern. The optical mask can be implemented using an amplitude mask that attenuates the incoming light \cite{asif2016flatcam} or a phase mask that modulates based on diffraction \cite{boominathan2020phlatcam}. Due to the large PSF size and limited angular response of the pixels\cite{boominathan2020phlatcam,antipa2018diffusercam}, the convolution follows a zero-padded boundary condition.
% Amplitude masks are easier to fabricate but have lower light efficiency due to attenuation. Phase masks have better light efficiency, but the fabrication process involves photo-lithography and is significantly more complicated. 
Existing works like \cite{boominathan2020phlatcam,asif2016flatcam,antipa2018diffusercam} use a similar model for thin 2D and 3D lensless imaging. 
These lensless imagers were developed for imaging applications, and their ability to perform optical encryption has not been fully explored. % due to the inherent flaws in their design, which we later point out.

FlatCam \cite{asif2016flatcam} is a lensless camera that places a separable coded amplitude mask above a bare sensor array to enable a thin and flat form-factor imaging device, which can simulate a conventional camera by reconstructing conventional images from coded measurements. DiffuserCam \cite{antipa2018diffusercam} and PhlatCam \cite{boominathan2020phlatcam} replace the coded amplitude mask from FlatCam \cite{asif2016flatcam} with a coded phase mask for improved light efficiency and reconstruction quality. Spectral DiffuserCam\cite{monakhova2020spectral} exploits the multiplexing ability of lensless imagers to do hyper-spectral imaging. \opencam is similar to these methods in that a coded optical mask is used to capture coded measurements. However, \opencam design goes a step further - our aim is not only to enable high-quality reconstruction of conventional images but also to prevent decryption attacks. To this end, we employ a second optical scaling mask positioned flush against the bare sensor array and propose novel mask designs for the multiplexing and scaling masks.

% \textbf{Double-mask Lensless Imaging.} Although there are different possible arrangements of masks in a double mask lensless camera, in this work, we consider a special co-axial arrangement - the first mask with pattern $S$ is placed flush against the sensor, while a second mask is placed at some distance from the sensor. If the PSF due to the second mask is given by $P$, then the measurement recorded at the sensor for a scene $X$ is given by,
% \begin{equation}
%     Y = S(P*X) + N.
% \end{equation}
% The effect of the first mask is multiplicative, while that of the second mask is convolutional. Since the first mask acts like a scaling mask, it is implemented using an amplitude mask. The convolutional mask can be implemented using either an amplitude or a phase mask as described previously for single-mask lensless imaging.

%%%%%%%%%%%%%%%%%%%%%%%%%%%%%%%%%%%%%%%%%%%%%%%%%%%%%%%%%%%%%%%%%%%%%%%%%%
\subsection{Image Reconstruction} 
%%%%%%%%%%%%%%%%%%%%%%%%%%%%%%%%%%%%%%%%%%%%%%%%%%%%%%%%%%%%%%%%%%%%%%%%%%

Image reconstruction is a core problem in computational imaging, and plays a key role in lensless imaging. The problem of scene reconstruction involves the computational recovery of the latent scene from a lensless capture. Conventional approaches for lensless image reconstruction are mostly based on regularized least squares \cite{duarte2008single,asif2016flatcam,antipa2018diffusercam,antipa2019video}. More recently, deep-learning-based reconstruction methods have proposed 
\cite{khan2020flatnet,bagadthey2022flatnet3d,pan2022image,rego2021robust}. We employ both regularized least squares and deep-learning-based reconstruction methods. %For regularized least squares, we develop a novel formulation that accounts for our scaling mask.

%%%%%%%%%%%%%%%%%%%%%%%%%%%%%%%%%%%%%%%%%%%%%%%%%%%%%%%%%%%%%%%%%%%%%%%%%%
\section{Attack Model}
%%%%%%%%%%%%%%%%%%%%%%%%%%%%%%%%%%%%%%%%%%%%%%%%%%%%%%%%%%%%%%%%%%%%%%%%%%

Since optical elements behave as linear operators \cite{wetzstein2020inference}, optical encryption cameras will always be susceptible to chosen-plaintext attacks in which an attacker uses a large set of ciphertext-plaintext pairs to estimate the encryption function. Naturally, this limits the practical use of optical encryption cameras to settings where attackers lack physical access to the camera location. Accordingly, in this paper, we assume that attackers do not have physical access to the camera locations, but can access a camera's stream. In other words, attackers have access to ciphertexts (measurements) from a camera, but not their corresponding plaintext (raw images), except for three special cases which we detail below. This is a reasonable assumption for many settings, such as security cameras in homes and offices. 

In this paper, we consider four types of attacks: ciphertext-only attacks (COA) and three special cases of known plaintext attacks, namely, impulse known plaintext attack (I-KPA), uniform known plaintext attack (U-KPA), and uniform-impulse known plaintext attack (UI-KPA). The reason we consider the known impulse/uniform plaintext settings is that bright impulse-like illumination sources and uniform backgrounds often appear naturally in scenes and may be recognizable in the corresponding encrypted sensor measurement. Hence, attackers may gain access to an approximate impulse or uniform response of the sensor, without having physical access to the sensor. 

Let $C: \mathbb{R}^{W_1 \times H_1 \times 3} \times \mathbb{R}^{W_2 \times H_2 \times M} \rightarrow \mathbb{R}^{W_3 \times H_3 \times 3}$ denote an optical encryption camera that maps a plaintext (scene) $X \in \mathbb{R}^{W_1 \times H_1 \times 3}$ and key $K \in \mathbb{R}^{W_2 \times H_2 \times M}$ to the ciphertext (measurement) $Y = C(K,X)$. The goal of an attacker is to recover the plaintext $X$ from ciphertext $Y$ with knowledge of function $C$, but partial or no knowledge of key $K$. 

For each of the four attack types, we train a Dense Prediction Transformer (DPT) \cite{ranftl2021vision} $D$ to learn the inverse mapping from $B=\{Y,\hat{X},\hat{K}\}$ for arbitrary key $K$ to the corresponding plaintext $X$, where $Y$ denotes a single cipherext, $\hat{X}$ an estimate of the plaintext, and $\hat{K}$ an estimate of the key. For the COA, no estimates $\hat{X}$ and $\hat{K}$ are available, so $B$ reduces to $Y$. For the KPAs, the methods for estimating $\hat{X}$ and $\hat{K}$ depend on the design of the targeted optical encryption camera (See section \ref{expt_sect} for additional details.)

To train $D$, we generate a random key for each ciphertext sample in each batch and use a loss consisting of a combination of an L1 pixel loss and an L2 perceptual loss (as in \cite{ledig2017photo,dosovitskiy2016generating}) over the outputs of layers \emph{relu1\_1}, \emph{relu2\_2}, and \emph{relu3\_3} of VGG16 \cite{simonyan2014very} pre-trained for image classification on the ImageNet \cite{deng2009imagenet} dataset. Concretely, the loss is given by 
\begin{equation}
\mathcal{L}_{D} = w_1||D(B)-X||_1 + w_2\sum_{i=1}^3 ||\phi_i(D(B))-\phi_i(X)||_2^2,
\label{eq:attack_loss}
\end{equation}
where $w_1=0.5$, $w_2=1.2$, and
$\phi_1: \mathbb{R}^{H\times W\times 3} \rightarrow \mathbb{R}^{\frac{H}{2}\times \frac{W}{2}\times 64}$,
$\phi_2: \mathbb{R}^{H\times W\times 3} \rightarrow \mathbb{R}^{\frac{H}{4}\times \frac{W}{4}\times 128}$, and
$\phi_3: \mathbb{R}^{H\times W\times 3} \rightarrow \mathbb{R}^{\frac{H}{8}\times \frac{W}{8}\times 256}$ denote the layers \emph{relu1\_1}, \emph{relu2\_2}, and \emph{relu3\_3}, respectively, of the pre-trained VGG16 network.

\begin{figure*}
    \centering
    \includegraphics[width=\linewidth]{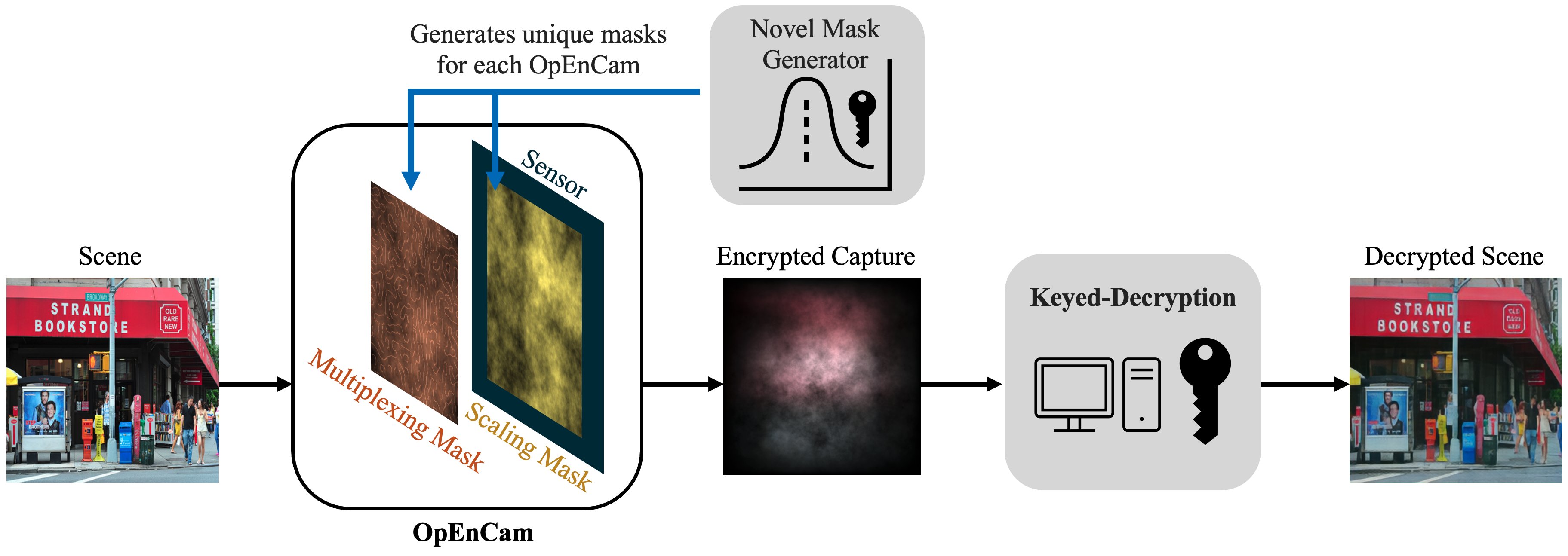}
    \caption{\textbf{\opencam Framework.} Our \textbf{Op}tical \textbf{En}cryption \textbf{Cam}eras encrypt image data before image capture using optical masks. Recovery of decrypted image data is possible only if you have access to the camera's encryption key, which is defined by the unique optical elements of each camera. The optical mask patterns are generated by a novel mask generator that is robust against various ciphertext-only attacks.}
    \label{fig:arch}
\end{figure*}

%%%%%%%%%%%%%%%%%%%%%%%%%%%%%%%%%%%%%%%%%%%%%%%%%%%%%%%%%%%%%%%%%%%%%%%%%%
\section{\opencam}
%%%%%%%%%%%%%%%%%%%%%%%%%%%%%%%%%%%%%%%%%%%%%%%%%%%%%%%%%%%%%%%%%%%%%%%%%%
% \opencam has three main components: (a) the double-masked lensless imaging model (encryption process), (b) mask design (key generation), and (b) image reconstruction (keyed decryption). Please see \cref{fig:arch}. 

% \opencam has three main components: (a) , (b) the mask design module (key generation), and (b) image reconstruction (keyed decryption). Please see \cref{fig:arch}. 

% The optics design module generates random optical masks to be used for constructing each optical encryption camera. The sensor module directly captures encrypted images via optical filtering of the incident light-field. The decryption module recovers all-in-focus images of the scene from encrypted measurements. The details of the three modules are illustrated in .  

%%%%%%%%%%%%%%%%%%%%%%%%%%%%%%%%%%%%%%%%%%%%%%%%%%%%%%%%%%%%%%%%%%%%%%%%%%
\subsection{Imaging Model (Encryption Process)}\label{section:encryption}
%%%%%%%%%%%%%%%%%%%%%%%%%%%%%%%%%%%%%%%%%%%%%%%%%%%%%%%%%%%%%%%%%%%%%%%%%%

We propose a novel double-mask sensor design with two co-axially placed masks - an optical scaling mask $S \in \mathbb{R}^{W_3 \times H_3 \times 3}$ positioned flush against the sensor array and an optical multiplexing mask positioned a few millimeters above the scaling mask. The scaling mask is implemented via an amplitude mask. For the multiplexing mask, an amplitude or phase mask can be used, as described in the previous section. Let $P \in \mathbb{R}^{W_2 \times H_2 \times 3}$ denote the point-spread-function (PSF) produced by the multiplexing mask. Then, $P$ and $S$ constitute the “encryption key” of the system, and the encrypted measurements $Y \in \mathbb{R}^{W_3 \times H_3 \times 3}$ captured by the sensor are given by
\begin{equation}
    Y = S \cdot (P \ast X) + N,
\end{equation}
where $X \in \mathbb{R}^{W_1 \times H_1 \times 3}$ denotes the scene and $N \in \mathbb{R}^{W_3 \times H_3 \times 3}$ the sensor noise. It should be noted again that $\ast$ represents a full-sized convolution with zero-padded boundary conditions similar to the model used in existing thin lensless cameras.

Compared to single-mask lensless designs \cite{liao2017ciphertext,boominathan2020phlatcam,ishii2020privacy,wang2019privacy,asif2016flatcam}, our novel double-mask design has a larger key space, which makes decryption attacks more challenging, and also inhibits impulse and uniform known plain text attacks. For the single mask design, the encryption key (i.e., PSF $P$) is equal to the impulse response of the camera, so the IKPA completely compromises the system. This is not the case for our double-mask design, as the scaling mask makes the system shift-variant, so the impulse response only reveals the ``effective`` encryption key for a single scene pixel. 

\vspace{1em}
\noindent \textbf{Imaging a uniform scene.} We refer to the response of the camera to a uniform scene as Uniform Scene Response (USR). Because of the structure of \opencam forward imaging model, it is natural to wonder if a USR, due to scenes like a plain wall, can reveal the scaling mask $S$ up to a scale. For a uniform scene to reveal the scaling mask $S$, $P*X$ must also be uniform. However, due to full-sized, linear convolution with zero-padded boundary conditions, $P*X$ is never uniform, even for a plain wall scene. Therefore, it is not possible to reveal $S$ from a USR.

\subsection{Mask Design (Key Generation)}\label{keygen}

For both single and double-mask optical encryption cameras, the design of the masks is critical for good performance, i.e., for enabling high-quality keyed decryption while also thwarting decryption attacks. Consider the design with a single multiplexing mask. The mask should have the following desirable properties:
\begin{enumerate}%[noitemsep,nolistsep,leftmargin=*]
    \item PSF should contain directional filters for all angles to enable high-quality lensless reconstruction \cite{boominathan2020phlatcam}. 
    \item The autocorrelation of the PSF should not be an impulse. The PSFs for which the autocorrelation is an impulse, are susceptible to autocorrelation-based COAs \cite{liao2017ciphertext}. These attacks exploit the fact that taking the autocorrelation of encoded measurements eliminates the mask component if the autocorrelation of the PSF is an impulse. This reduces the COA problem to recovering an image from its autocorrelation.
    \item The PSF produced by the multiplexing mask should not be binary. Binary PSFs may be revealed to an attacker when a point source appears in the scene. Although our double-mask design inherently provides protection against such attacks, it is still not desirable for the PSF to be compromised, as it reduces the key size significantly. Consider a 1D case of binary PSF $p$, and positive scaling mask $s$. The measurement $y(n)$ recorded at the sensor due to a single point source is given by,
\begin{equation}
    y(n) = 
    \begin{cases}
        s(n),& \text{if } n \in \mathcal{J}\\
        0,              & \text{otherwise},
    \end{cases}
\end{equation}
where $\mathcal{J}$ is the set of all pixel locations for which $p(n)$ is 1.
Simple thresholding of $y$ i.e. $y > 0$, reveals $p$ and reduces the size of the key.
\end{enumerate}
Keeping the above criteria in mind, we propose a novel design for the multiplexing mask. The PSF due to \opencam multiplexing mask is given by,
\begin{equation}
    P = \alpha P_{colr}  + (1-\alpha)P_{cont},
\end{equation}
where $P_{cont}$ is a binary contour PSF obtained from Perlin noise\cite{perlin2002improving} proposed in \cite{boominathan2020phlatcam}, $P_{colr}$ is a colored noise with certain roll-off. Perlin contours are known for high-quality lensless imaging as shown in \cite{boominathan2020phlatcam,khan2020flatnet} but have impulse-like autocorrelations. On the hand, $P_{colr}$ is smoother and has non-impulse autocorrelation with varying roll-off. We fix the Perlin feature size, randomize the permutation vector for Perlin noise, and correspondingly the contour PSF $P_{cont}$. The length of the permutation vector is the same as the height or width of the PSF. To generate $P_{colr}$, we generate colored noise with Power Spectral Density (PSD) given by,
\begin{equation}
    H_{\beta}(u,v) = \frac{1}{(u^2+v^2)^{(\frac{\beta}{2})}}.
\end{equation}
The corresponding colored noise $P_{colr}(\beta)$ is given by,
\begin{equation}
    P_{colr}(\beta) = \mathcal{F}^{-1}(\sqrt{H_{\beta}}e^{j\theta}).
\end{equation}
Here $\mathcal{F}^{-1}$ is the inverse FFT, $\theta$ is random white noise sampled from $\mathcal{U}(0,1)$ and $\beta$ is a scalar hyper-parameter that controls the roll-off of the colored noise. We randomly sample $\beta$ from $\mathcal{U}(1,10)$. Finally, the linear combination co-efficient $\alpha$ is chosen uniformly at random. Apart from the above criteria, it should also be noted that PSF must be non-negative as we can not subtract light. 

\begin{figure}[t]
    \centering
    \includegraphics[width=\linewidth]{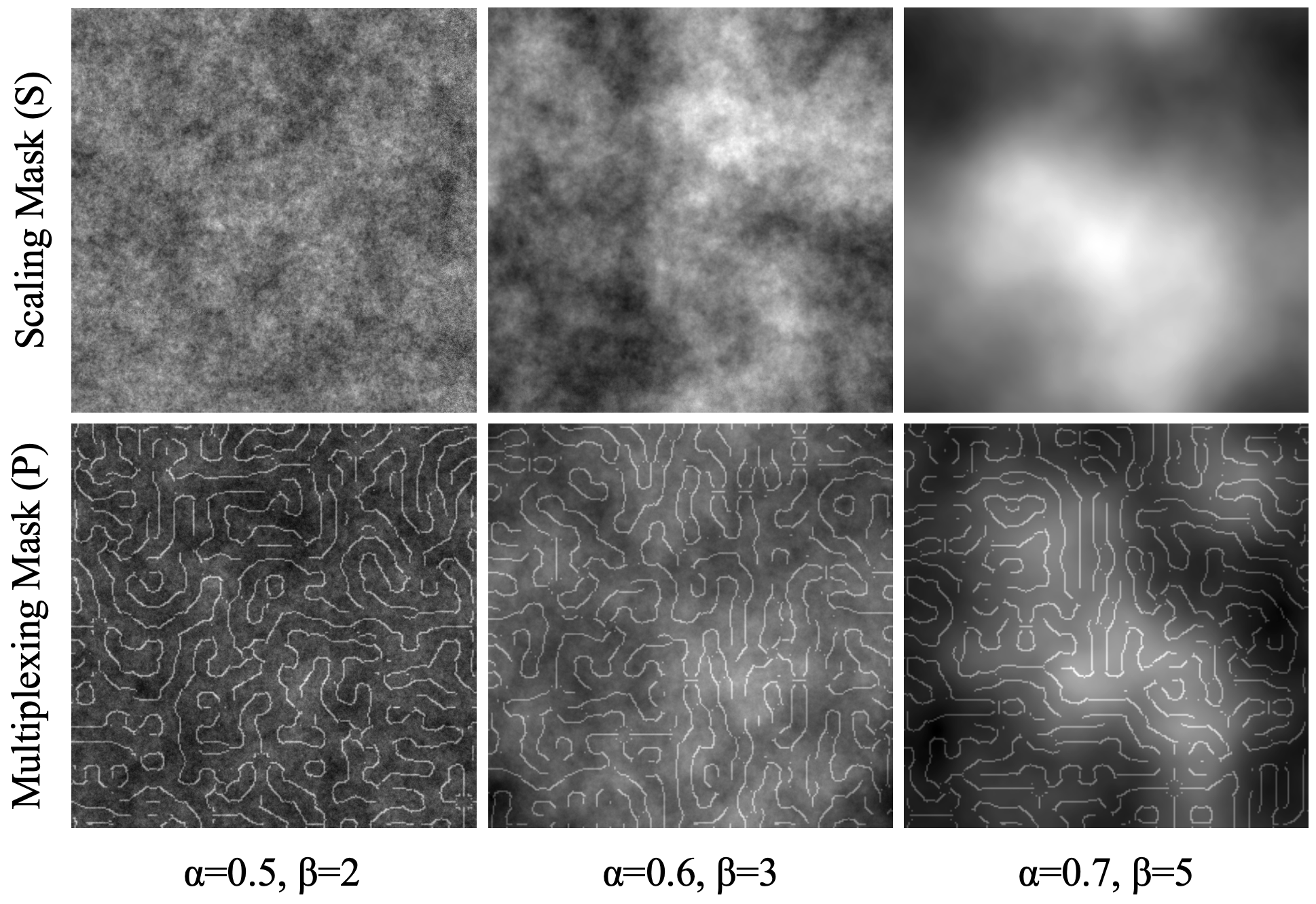}
    \caption{\textbf{Generated Mask Samples.} Each column shows a scaling mask and multiplexing mask pair generated from our key generator. Each of these is randomly generated. Moreover, the underlying Perlin contours in the multiplexing mask PSF are also randomized.}
    \label{fig:mask_patterns}
\end{figure}

For the scaling mask $S$, we use colored noise again without the Perlin contours. For a given pair of $(S,P)$, the color of the noise ($\beta$) is the same for both $S$ and $P_{colr}$ to avoid attacks due to filtering of the either component. It should be noted that $S$ has to be positive as we do not want to throw information from any of the sensor pixels. Some samples of the generated mask patterns are shown in \cref{fig:mask_patterns}.

\begin{figure}[t]
    \centering
    \includegraphics[width=\linewidth]{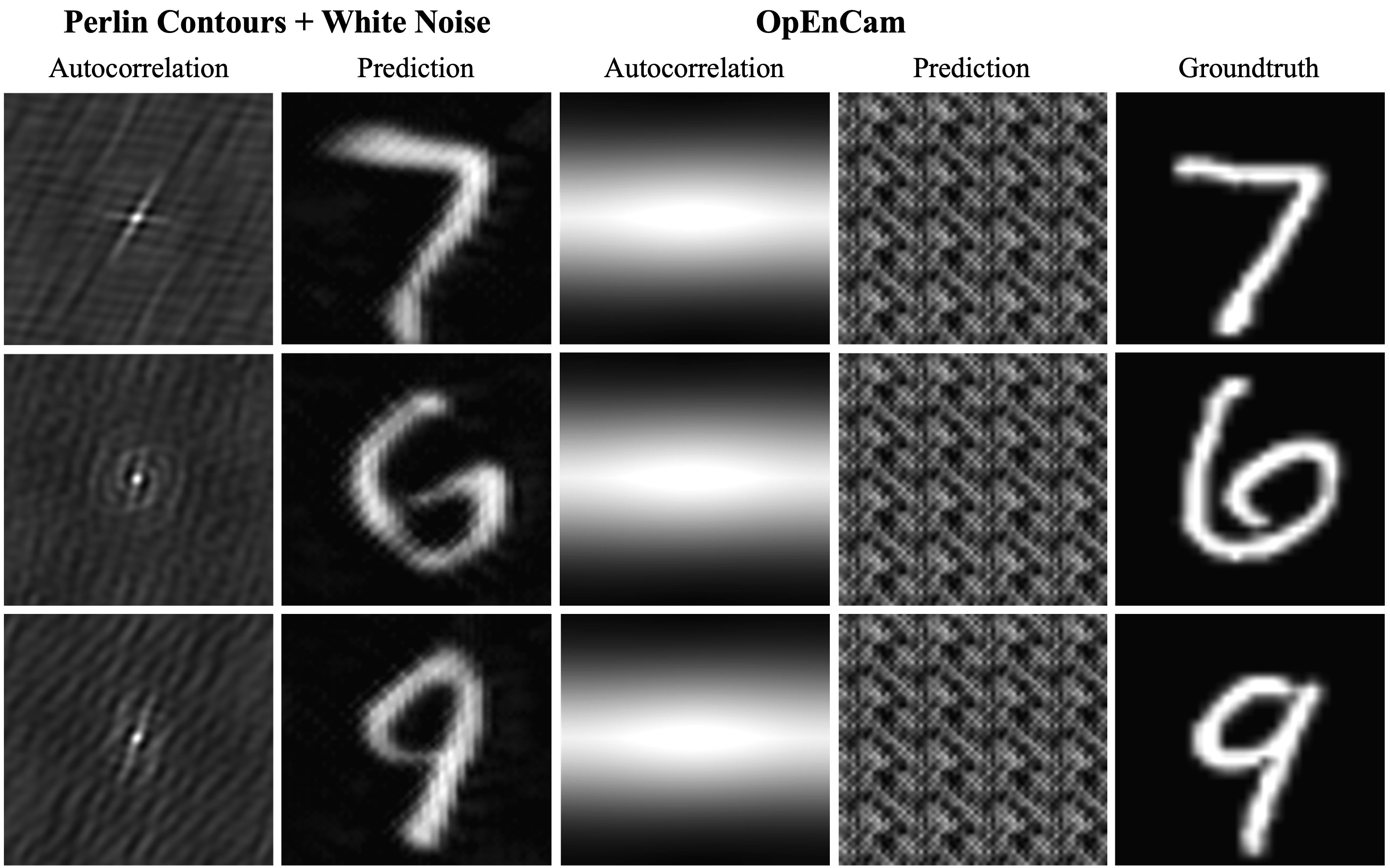}
    \caption{\textbf{Autocorrelation Analysis.} It is not possible to reconstruct the underlying plaintext/scene from the measurement autocorrelation for the \opencam design $P_O$. While the reconstruction is possible for $P_W$ (Perlin Contours + White noise).}
    \label{fig:autocorr_at}
\end{figure}

\subsection{Effect of Colored Noise PSF - Autocorrelation Analysis}\label{autocorr}
In the above subsection, we claimed that autocorrelation of the multiplexing mask should not be an impulse and, as a result, settled on a combination of Perlin contour and colored noise for the multiplexing PSF of \opencam. In this experiment, we validate this claim by comparing the performance of two different multiplexing mask designs under autocorrelation-based ciphertext-only attack. The mask designs being compared are (a) $P_O = \alpha P_{colr}  + (1-\alpha)P_{cont}$, (b)$P_W = \alpha P_{white}  + (1-\alpha)P_{cont}$, where $P_{white}$ is white noise PSF with impulse autocorrelation. $P_O$ is the \opencam design while $P_W$ is a corresponding version with white noise used instead of colored noise for PSF. To perform this experiment, we first simulate measurements using each of the mask generators and perform autocorrelation of the simulated measurements. We then feed this autocorrelation to a U-Net, which then learns to estimate the underlying scene or plaintext from the autocorrelation. Given, that the autocorrelation of $P_W$ is close to an impulse (because autocorrelations of $P_{white}$ and $P_{cont}$ are impulse), U-Net finds it easy to learn the mapping from the measurement autocorrelation to the underlying scene for $P_W$, while it fails to do so for $P_O$. We perform this experiment on MNIST dataset. Fig. \ref{fig:autocorr_at} shows the visual results for the experiment. As can be seen, it is possible to reconstruct the underlying scene from the autocorrelation of the measurement for $P_W$, while this is not the case for $P_O$ i.e. our \opencam design.

\begin{figure*}
    \centering
    \includegraphics[width=\linewidth]{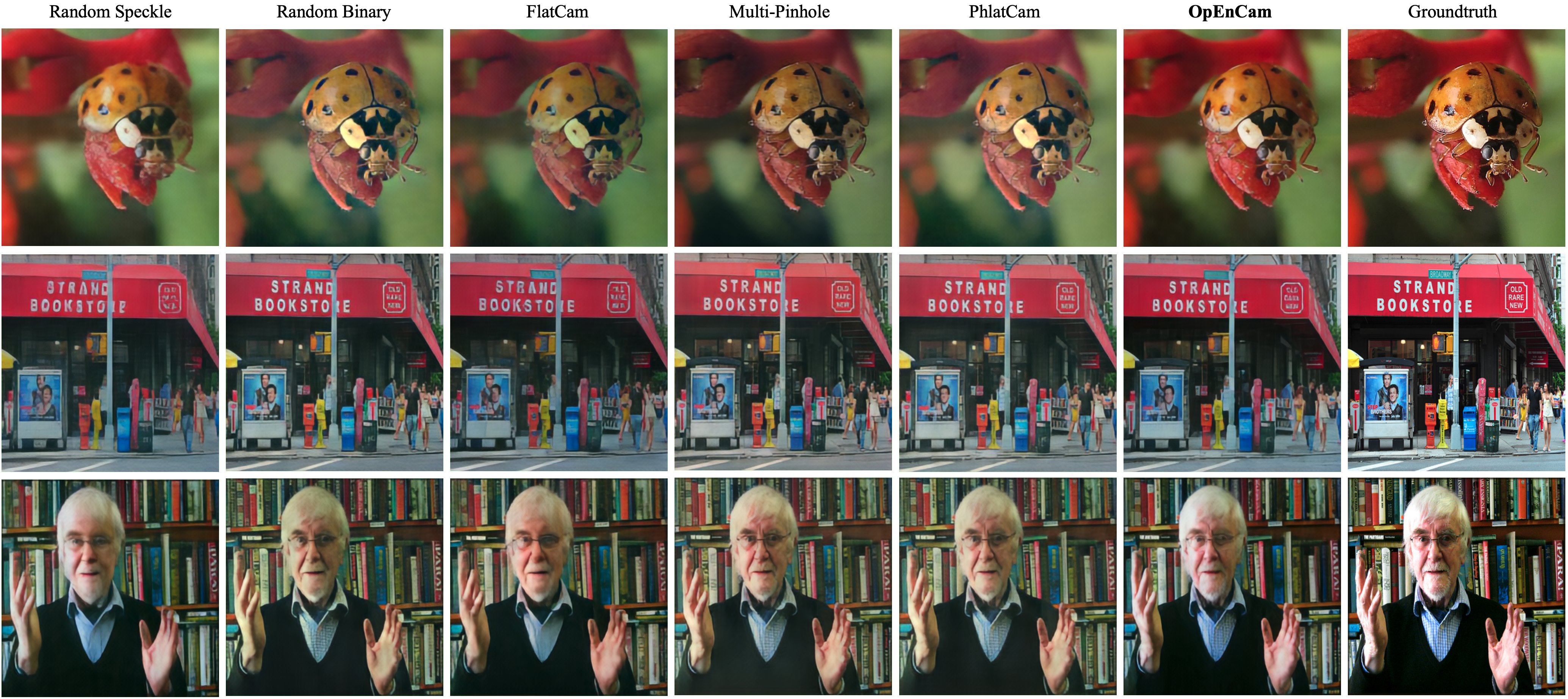}
    \caption{\textbf{Keyed Decryption.} The first five columns show the keyed-decryption performance for existing single mask systems. Keyed decryption using PhlatCam (Boominathan \textit{et al.}\cite{boominathan2020phlatcam}) is currently state of the art for lensless imaging. \opencam keyed decryption performance matches that of existing mask designs for lensless imaging and optical encryption despite the addition of a scaling mask to enhance privacy.}
    \label{fig:tf_keyed}
\end{figure*}

\begin{table}[t]
\begin{center}
\centerline{\resizebox{\columnwidth}{!}{
\begin{tblr}{c|ccc}
\toprule
\textbf{Encryption Design} & {\textbf{PSNR} $\uparrow$} &\textbf{SSIM} $\uparrow$ & \textbf{LPIPS} $\downarrow$\\
\midrule
 PhlatCam \cite{boominathan2020phlatcam} & \textbf{23.10} & \textbf{0.81} & \textbf{0.30}\\
 Multi-pinhole \cite{ishii2020privacy} & 22.70 & 0.79 & \textbf{0.30}\\
 FlatCam \cite{nguyen2019deep} & 20.68 & 0.74 & 0.40\\
 Random Binary \cite{wang2019privacy} & 20.22 & 0.77 & 0.35\\
 Random Speckle \cite{Zang:13} & 20.43 & 0.71 & 0.44\\
\opencam & \underline{22.90} & \underline{0.80} & \underline{0.32}\\
\bottomrule
  \end{tblr}
}}
\end{center}
\caption{\textbf{Image Recovery via Keyed Decryption.} The reconstruction quality of \textbf{OpEnCam} is at par with state of the art lensless image reconstruction quality. Best performance is made bold while the second best performance is underlined.}
\label{tab:recon}
\end{table}

\subsection{Image Reconstruction (Keyed Decryption)}\label{keyedec}

The decryption module accepts three inputs, an encrypted image $Y$, a scaling mask $S$, and a PSF $P$ due to the multiplexing mask $M$; and returns a decrypted image $\hat{X}$. For an ideal system, decryption can be achieved by simply applying the inverse of the two optical computations. However, in practice, to account for noise, cropping, and imperfect calibration, we instead employ a double-step estimation process to recover the scene $X$. In the first step, we employ Tikhonov regularized least squares to recover an initial estimate $ \hat{X_{T}}$ of image $X$:
% \begin{equation}\label{tikhonov}
%     \hat{X}_T = \argmin_X || Y -  S \cdot (M \ast X) ||^2_2 + ||X||^2_2
% \end{equation}
\begin{equation}\label{tikhonov}
    \hat{X}_T = \argmin_X || Y_{N} -  P \ast X ||^2_F + \gamma||X||^2_F
\end{equation}
Here $Y_{N}$ is the measurement after scaling normalization i.e. $Y_{N} = \frac{Y}{S+\epsilon}$. $\hat{X}_T$ has a closed-form solution and can be implemented in the Fourier domain as Wiener filtering. 

In the second step, we feed our initial estimate $\hat{X_{T}}$ plus measurement $Y$, and multiplexing mask PSF $P$ to a Dense Prediction Transformer (DPT) \cite{ranftl2021vision}  $D: \mathbb{R}^{W_1\times H_1 \times 7} \rightarrow \mathbb{R}^{W_1\times H_1 \times 3}$, which we train via stochastic gradient descent to produce a refined version $\hat{X}$, given input $B=\{Y,\hat{X_T},P\}$. As in \cref{eq:attack_loss}, we use a combination of L1 loss and VGG feature loss to learn the mapping with $w_1=0.5$ and $w_2=1.2$. We found that this transformer-based refinement performed slightly better than the U-Net-based refinement typically used in lensless imaging works like \cite{khan2019towards,khan2020flatnet,monakhova2019learned}. 

% \end{enumerate}

%%%%%%%%%%%%%%%%%%%%%%%%%%%%%%%%%%%%%%%%%%%%%%%%%%%%%%%%%%%%%%%%%%%%%%%%%%
\section{Experimental Results}\label{expt_sect}
%%%%%%%%%%%%%%%%%%%%%%%%%%%%%%%%%%%%%%%%%%%%%%%%%%%%%%%%%%%%%%%%%%%%%%%%%%

\subsection{Preliminaries}

\subsubsection{Baselines} 
We compare the performance of \opencam design with other mask-based passive optical encryption designs as well as existing lensless camera designs. Our baselines for optical encryption include multi-pinhole mask generator (Ishii \textit{et al.}\cite{ishii2020privacy}), random binary mask generator (Wang \textit{et al.}\cite{wang2019privacy}) and random speckle mask generator (Zang \textit{et al.}\cite{Zang:13}) We also compare against PhlatCam\cite{boominathan2020phlatcam} lensless camera mask generator, which uses a Perlin contour PSF multiplexing mask, and FlatCam\cite{asif2016flatcam,nguyen2019deep}, which uses separable Maximum Length Sequences (MLS) for mask patterns. 

\subsubsection{Dataset and Implementation}
We validate the efficacy of our proposed framework via simulation and real-world experiments. For training keyed-decryption and attack models described in this section, we use the Places365 dataset \cite{zhou2017places}. We use a subset of ImageNet\cite{ILSVRC15} for quantitative evaluation of the keyed-decryption and attack models. Given that each encryption approach (including \opencam) has its own mask generator that randomly generates the corresponding mask patterns, we simulate the captured measurement/ciphertext using these randomly generated mask patterns using the above datasets. Once simulated, we feed these measurements (along with the mask patterns for keyed-decryption) to the corresponding neural networks for final predictions. During testing, we generate predictions for a large number of mask patterns generated from each mask generator and report the average performance.

% Apart from the existing works, we also compare against three additional naive optical encryption designs that we propose - (a) speckle with scaling mask (Speckle Double Mask), (b) single colored noise multiplexing mask (Colored Single Mask), and (c) colored noise multiplexing mask with a scaling mask (Colored Double Mask).

\begin{table}[t]
\begin{center}
\centerline{\resizebox{\columnwidth}{!}{
\begin{tblr}{c|ccc}
  \toprule
\textbf{Encryption Design} & {\textbf{PSNR} $\downarrow$} &\textbf{SSIM} $\downarrow$ & \textbf{LPIPS} $\uparrow$\\
\midrule
 PhlatCam \cite{boominathan2020phlatcam} & \underline{18.10} & \underline{0.63} & \underline{0.57}\\
 Multi-pinhole \cite{ishii2020privacy} & 19.29 & 0.65 & 0.48\\
 FlatCam \cite{nguyen2019deep} & 19.00 & 0.65 & 0.54\\
 Random Binary \cite{wang2019privacy} & 18.73 & 0.64 & 0.56\\
 Random Speckle \cite{Zang:13} & 19.42 & 0.65 & 0.54\\
\opencam - Single Mask & 18.22 & 0.64 & \underline{0.57}\\
\opencam - Double Mask & \textbf{16.28} & \textbf{0.59} & \textbf{0.60}\\
  \bottomrule
  \end{tblr}
}}
\end{center}
\caption{\textbf{Transformer-based Blind COA.} Double mask \textbf{OpEnCam} provides the best security as seen from the reconstruction quality. Large gap between single and double mask \opencam highlights the improvement offered by the scaling mask in terms of privacy. Best performance is made bold while the second best performance is underlined.}
\label{tab:blind}
\end{table}

%%%%%%%%%%%%%%%%%%%%%%%%%%%%%%%%%%%%%%%%%%%%%%%%%%%%%%%%%%%%%%%%%%%%%%%%%%
\subsection{Keyed Decryption}\label{keyed_expt}
%%%%%%%%%%%%%%%%%%%%%%%%%%%%%%%%%%%%%%%%%%%%%%%%%%%%%%%%%%%%%%%%%%%%%%%%%%

% \begin{table}[t]
% \centering
% \caption{\textbf{Image Recovery via Keyed Decryption.} The reconstruction quality of \textbf{OpEnCam} is at par with state of the art lensless image reconstruction quality. Best performance is made bold while the second best performance is underlined.}
% \resizebox{\columnwidth}{!}{
% \begin{tabular}{|c|c|c|c|}
% \hline
% \textbf{Method} & {\textbf{PSNR} $\uparrow$} &\textbf{SSIM} $\uparrow$ & \textbf{LPIPS} $\downarrow$\\
% \hline
%  \textbf{PhlatCam}\cite{boominathan2020phlatcam} & \textbf{23.10} & \textbf{0.81} & \textbf{0.30}\\
% \hline
%  Multi-pinhole\cite{ishii2020privacy} & 22.70 & 0.79 & 0.30\\
% \hline
%  FlatCam\cite{nguyen2019deep} & 20.68 & 0.74 & 0.40\\
% \hline
%  Random Binary\cite{wang2019privacy} & 20.22 & 0.77 & 0.35\\
% \hline
%  Random Speckle\cite{Zang:13} & 20.43 & 0.71 & 0.44\\
% \hline
% \underline{\opencam} & \underline{22.90} & \underline{0.80} & \underline{0.32}\\
% \hline
% \end{tabular}
% }
% \label{tab:recon}
% \end{table}

Given that high-quality scene recovery using the key or mask patterns defines the usefulness of our system, we first experimentally validate the performance of keyed decryption transformer. We compare the keyed-decryption performance of \opencam against PhlatCam \cite{boominathan2020phlatcam}, Multi-pinhole \cite{ishii2020privacy}, FlatCam \cite{nguyen2019deep}, Random Binary \cite{wang2019privacy} and Random Speckle \cite{Zang:13}. We use the same keyed-decryption strategy as described in Section \ref{keyedec} for a fair comparison. The keyed-decryption network uses Tikhonov reconstructions described in Equation \ref{tikhonov} along with the measurement and the multiplexing mask as input and produces the final reconstructions. We use DPT \cite{ranftl2021vision} as the transformer. We compute the average Peak Signal to Noise ratio (PSNR), Structural Similarity Index Measure (SSIM) and Learned Perceptual Image Patch Similarity (LPIPS) for a subset of ImageNet dataset as described in \cite{khan2020flatnet}. Table \ref{tab:recon} shows the average performance. Higher PSNR, SSIM and lower LPIPS indicate better reconstruction quality. The performance of \opencam is at par with state of the art lensless image reconstruction performance \cite{boominathan2020phlatcam,khan2019towards} while marginally outperforming existing encryption methods like \cite{ishii2020privacy,wang2019privacy,Zang:13,nguyen2019deep}. A similar trend is observed in visual results shown in \cref{fig:tf_keyed}.

%%%%%%%%%%%%%%%%%%%%%%%%%%%%%%%%%%%%%%%%%%%%%%%%%%%%%%%%%%%%%%%%%%%%%%%%%%
\subsection{COA: Ciphertext Only Attack}\label{coa_expt}
%%%%%%%%%%%%%%%%%%%%%%%%%%%%%%%%%%%%%%%%%%%%%%%%%%%%%%%%%%%%%%%%%%%%%%%%%%

\begin{figure*}
    \centering
    \includegraphics[width=\linewidth]{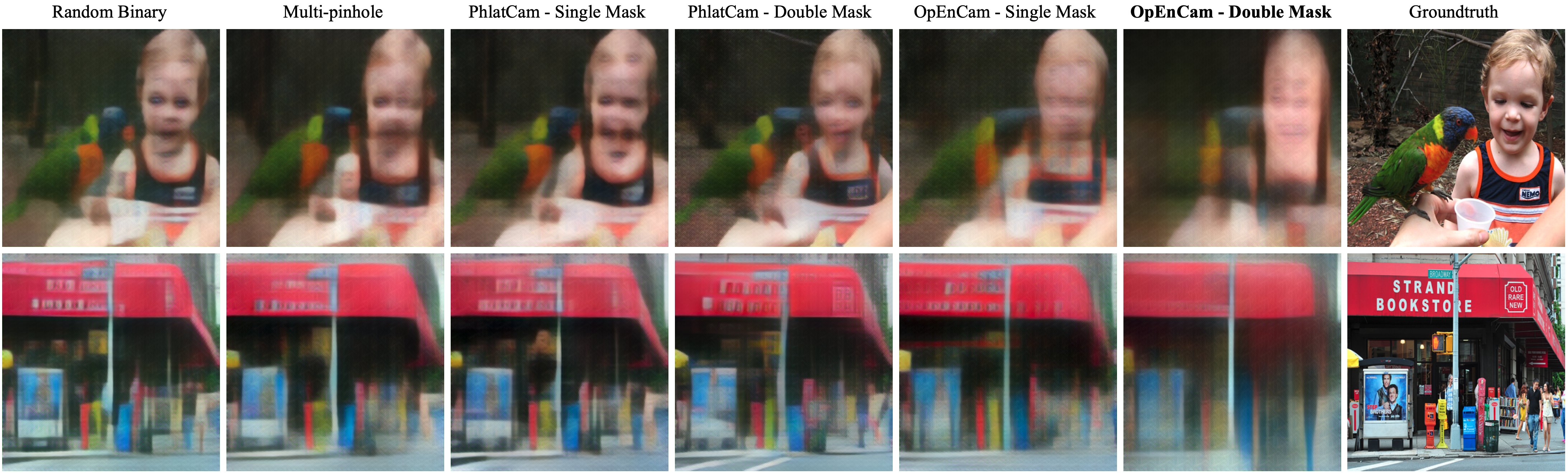}
    \caption{\textbf{Blind COA.}We show some visual reconstruction results for two scenes. Double-mask \opencam reconstructions are barely recognizable, while existing approaches reveal a significant amount of texture.}
    \label{fig:tf_blind}
\end{figure*}

We train a Dense Prediction Transformer(DPT) \cite{ranftl2021vision} to learn a mapping from a single ciphertext/measurement to the corresponding plaintext/scene for each of the mask generators using the loss function described in \ref{eq:attack_loss}. A secure optical encryption system should be robust against these powerful data-driven transformer-based attacks. We compare the scene reconstruction performance for different optical encryption strategies. Among the existing works, we compare against PhlatCam\cite{boominathan2020phlatcam}, random speckle\cite{Zang:13}, multi-pinhole\cite{ishii2020privacy}, random binary\cite{wang2019privacy} and FlatCam\cite{nguyen2019deep}. We also compare against single-mask \opencam i.e. the \opencam design without the scaling mask to highlight the importance of the double-mask system. Lower PSNR, SSIM and higher LPIPS indicate lower reconstruction quality and better security. In \cref{fig:priv_util_tci}, we show the privacy-utility tradeoff plot. From the plot, it can be observed that \opencam clearly outperforms existing optical encryption and lensless imaging designs. The reconstructions from \opencam, shown in \cref{fig:tf_blind}, are barely recognizable, indicating that it is extremely challenging to recover the original scene from a single encoded measurement captured using \opencam when the key or mask patterns are not available. Table \ref{tab:blind} shows the quantitative comparison of the different encryption approaches. Double Mask \opencam clearly outperforms all other existing optical encryption and lensless imaging designs. The large gap between \opencam single and double mask models suggests that the scaling mask not only makes the system shift-variant but also adds another layer of privacy to existing mask-based lensless camera designs.

\begin{table}[t]
\begin{center}
\centerline{\resizebox{\columnwidth}{!}{
\begin{tblr}{c|ccc}
  \toprule
\textbf{Encryption Design} & {\textbf{PSNR} $\downarrow$} &\textbf{SSIM} $\downarrow$ & \textbf{LPIPS} $\uparrow$\\
\midrule
 PhlatCam \cite{boominathan2020phlatcam} & 19.04 & 0.76 & 0.38\\
 Multi-pinhole \cite{ishii2020privacy} & 22.61 & 0.85 & 0.22\\
 FlatCam \cite{nguyen2019deep} & 18.79 & 0.66 & 0.52\\
 Random Binary \cite{wang2019privacy} & 19.28 & 0.72 & 0.43\\
 Random Speckle \cite{Zang:13} & 19.04 & 0.65 & 0.53\\
PhlatCam - Double Mask  & 18.28 & 0.67 & 0.50\\
\opencam - Single Mask & \underline{18.02} & \underline{0.66} & \underline{0.57}\\
\opencam - Double Mask & \textbf{16.28} & \textbf{0.59} & \textbf{0.60}\\
  \bottomrule
  \end{tblr}
}}
\end{center}
\caption{\textbf{Impulse - KPA.} Attacker has access to an approximate PSF $P$ corresponding to a single bright source. Double-mask \opencam outperforms all existing designs. PhlatCam even with a scaling mask (PhlatCam-Double) is susceptible to this attack due to its sparse binary PSF. Best performance is made bold while the second best performance is underlined.}
\label{tab:impulse}
\end{table}

% \begin{table}[t]
% \centering
% \caption{\textbf{Transformer-based Blind COA.} Double mask \textbf{OpEnCam} provides the best security as seen from the reconstruction quality. Best performance is made bold while the second best performance is underlined.}
% \resizebox{\columnwidth}{!}{
% \begin{tabular}{|c|c|c|c|}
% \hline
% \textbf{Method} & {\textbf{PSNR} $\downarrow$} &\textbf{SSIM} $\downarrow$ & \textbf{LPIPS} $\uparrow$\\
% \hline
%  \underline{PhlatCam}\cite{boominathan2020phlatcam} & \underline{18.10} & \underline{0.63} & \underline{0.57}\\
% \hline
%  Multi-pinhole\cite{ishii2020privacy} & 19.29 & 0.65 & 0.48\\
% \hline
%  FlatCam\cite{nguyen2019deep} & 19.00 & 0.65 & 0.54\\
% \hline
%  Random Binary\cite{wang2019privacy} & 18.73 & 0.64 & 0.56\\
% \hline
%  Random Speckle\cite{Zang:13} & 19.42 & 0.65 & 0.54\\
% \hline
% \opencam - Single Mask & 18.22 & 0.64 & \underline{0.57}\\
% \hline
% \textbf{\opencam - Double Mask} & \textbf{16.28} & \textbf{0.59} & \textbf{0.60}\\
% \hline
% \end{tabular}
% }
% \label{tab:recon}
% \end{table}

% \subsection{Cryptanalysis of OpEnCam}

%%%%%%%%%%%%%%%%%%%%%%%%%%%%%%%%%%%%%%%%%%%%%%%%%%%%%%%%%%%%%%%%%%%%%%%%%%
\subsection{I-KPA: Impulse - Known Plaintext Attack}
%%%%%%%%%%%%%%%%%%%%%%%%%%%%%%%%%%%%%%%%%%%%%%%%%%%%%%%%%%%%%%%%%%%%%%%%%%

\begin{figure*}
    \centering
    \includegraphics[width=\linewidth]{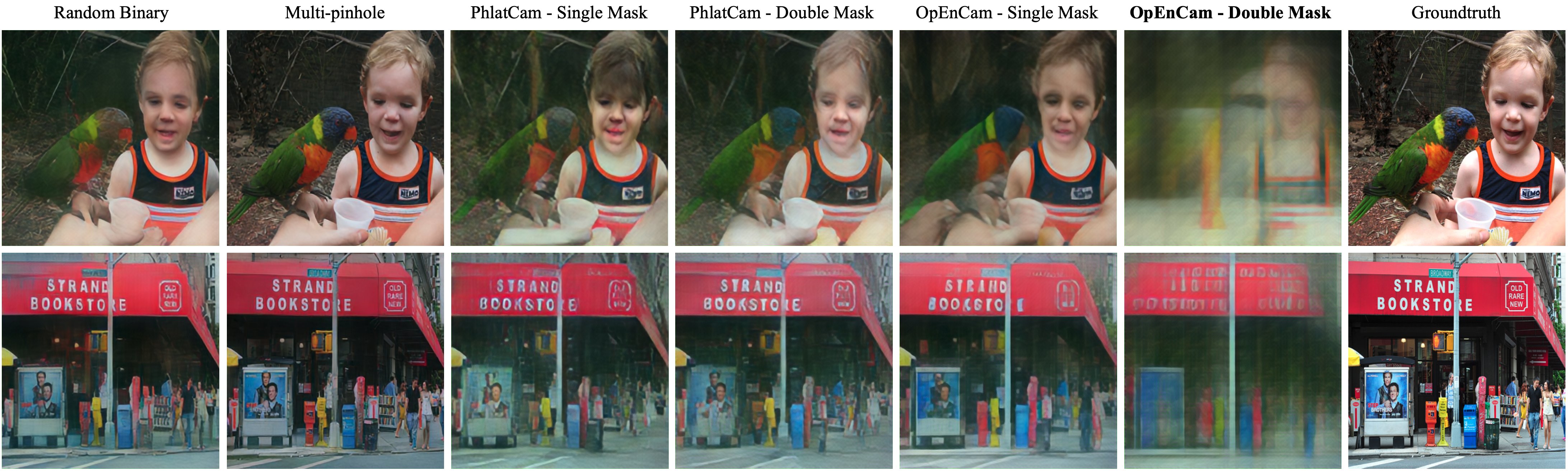}
    \caption{\textbf{Impulse - KPA.} We show some visual reconstruction results for two scenes. Double-mask \opencam reconstructions are barely recognizable, while other designs are completely susceptible to this attack, including Double-masked PhlatCam.}
    \label{fig:bs_attack_new}
\end{figure*}

\begin{figure}
    \centering
    \includegraphics[width=\linewidth]{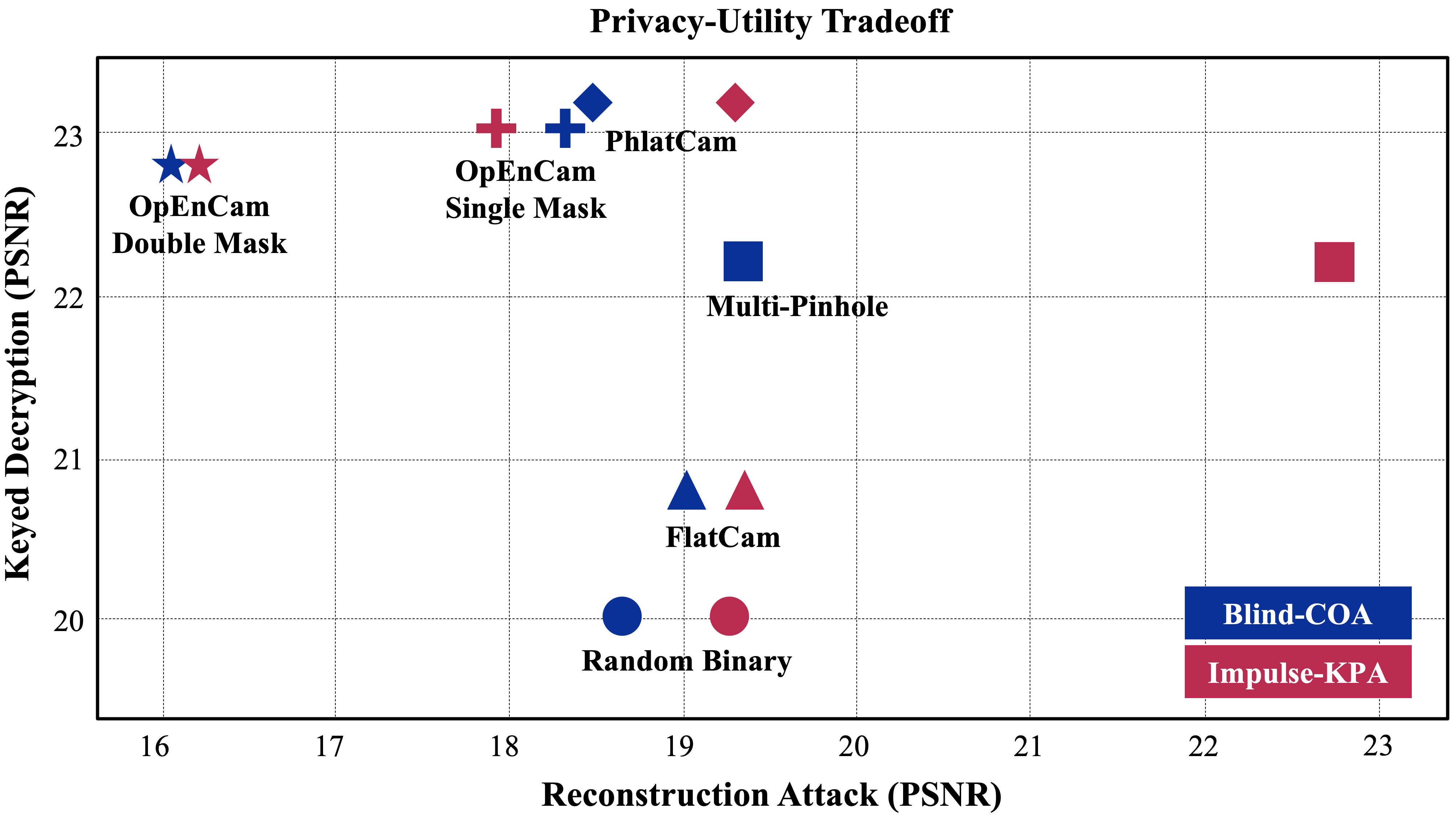}
    \caption{\textbf{Privacy-Utility Plots.} We show the privacy-utility plot for blind COA and Impulse-KPA. For both attacks, double-masked \textsc{OpEnCam}, outperforms other designs and occupies the desired top-left corner.}
    \label{fig:priv_util_tci}
\end{figure}

An optical encryption camera with a single multiplexing mask is susceptible to an Impulse based Known Plaintext Attack (I-KPA). More specifically, if a bright source appears in a scene, an approximate point spread function (PSF) due to the multiplexing mask is revealed. Since for single mask-based encryption, the PSF is the key, the attacker can use a bright-source measurement to decrypt ciphertexts.  Moreover, as pointed out in Section \ref{keygen}, binary masks are more susceptible to these bright source attacks even with a scaling mask. In this experiment, we show that our double-mask encryption of \opencam is robust against these bright source attacks. To perform this experiment, we first simulate a measurement of a scene with a bright source. The relative intensity of the bright source with respect to the rest of the scene is $10^3$. Assuming this measurement as the approximate PSF, we apply our keyed decryption pipeline of Section \ref{keyedec}. We compare against the baselines described above: PhlatCam \cite{boominathan2020phlatcam}, multi-pinhole \cite{ishii2020privacy}, random binary \cite{wang2019privacy}, FlatCam MLS \cite{nguyen2019deep}. To highlight the importance of our double-mask system we also compare it against single-mask version of \opencam without the scaling mask. Furthermore, to highlight the disadvantage of binary masks like that of PhlatCam under point source attacks, we compare against a modified version of PhlatCam where we add an additional scaling mask. We show the privacy-utility tradeoff plot for the I-KPA in \cref{fig:priv_util_tci}. Visual reconstruction results are shown in \cref{fig:bs_attack_new}. It can be seen that \opencam with scaling mask significantly outperforms all existing mask approaches. Moreover, despite the use of scaling mask, PhlatCam-Double Mask is unable to provide any privacy due to the binary nature of its multiplexing mask. Corresponding visual results for a few important approaches are also compared in the same figure. We show the quantitative results for the I-KPA in \cref{tab:impulse}. 

\subsection{U-KPA: Uniform - Known Plaintext Attack}
%%%%%%%%%%%%%%%%%%%%%%%%%%%%%%%%%%%%%%%%%%%%%%%%%%%%%%%%%%%%%%%%%%%%%%%%%%

\subsubsection{Attack due to uniform scene} 
\begin{figure}
    \centering
    \includegraphics[width=\linewidth]{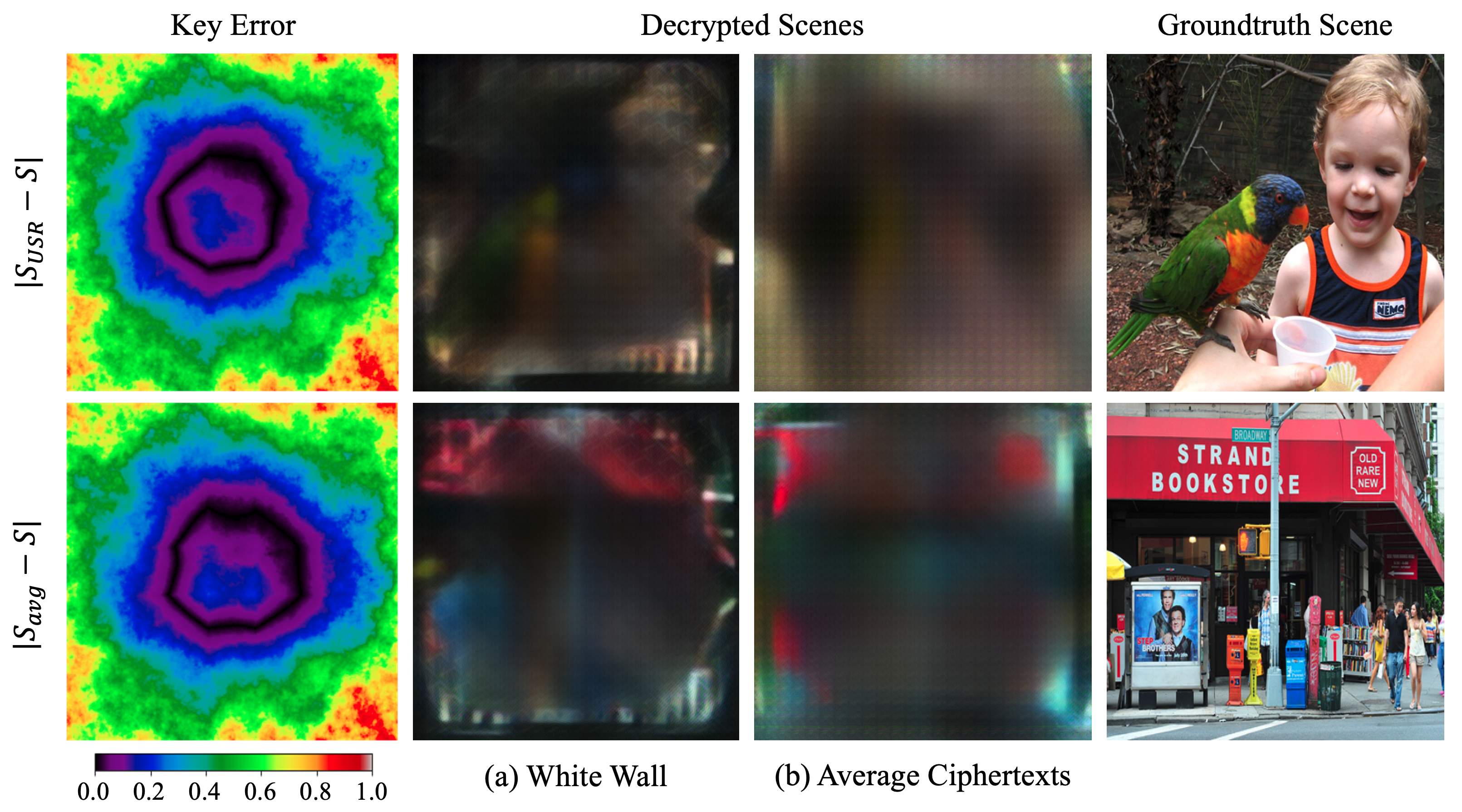}
    % \vspace{-.5cm}
    \caption{\textbf{Uniform - KPA.} (a) From a USR, (b) from ciphertexts. It can be seen that both the USR and the average of measurements fail to reveal the underlying hidden scaling mask leading to poor plain text estimations. } 
    % \vspace{-1.5em}
    \label{fig:avg_attack}
\end{figure}

To verify that a uniform scene response (USR) doesn't reveal the scaling mask, we approximated the scaling mask $S$ as $S_{USR} = Y_{USR}$, where $Y_{USR}$ is a measurement for an all-ones scene. We used $S_{USR}$ to remove the scaling mask component from the test measurement and decrypted it using the blind COA transformer trained for single-mask \opencam. 
We show the decrypted result and the error $|S_{USR} - S|$ in \cref{fig:avg_attack}(a). It can be seen that $S_{USR}$ is not an accurate approximation of $S$. 

\subsubsection{Attack due to multiple measurements}
Another question that may arise is whether an average of diverse measurements reveal the scaling mask. To verify this, we assume access to $N=12000$ diverse \opencam measurements of scenes from the Places365 dataset and estimate the scaling mask $S$ as $S_{avg} = (1/N)\sum_{j=1}^{N}Y^{(j)}$. We then used $S_{avg}$ to remove the scaling mask component from the measurement and decrypted it using the transformer trained for single-mask \opencam. 
We show the results in \cref{fig:avg_attack}(b). It can be seen that $S_{avg}$ is not an accurate approximation of $S$. This attack assumes that under the condition $N\rightarrow \infty$, $S_{avg}$ will be approximately a USR.

%%%%%%%%%%%%%%%%%%%%%%%%%%%%%%%%%%%%%%%%%%%%%%%%%%%%%%%%%%%%%%%%%%%%%%%%%%
\subsection{UI-KPA:  Uniform-Impulse Known Plaintext Attack}
%%%%%%%%%%%%%%%%%%%%%%%%%%%%%%%%%%%%%%%%%%%%%%%%%%%%%%%%%%%%%%%%%%%%%%%%%%

\begin{figure*}
    \centering
    \includegraphics[width=\linewidth]{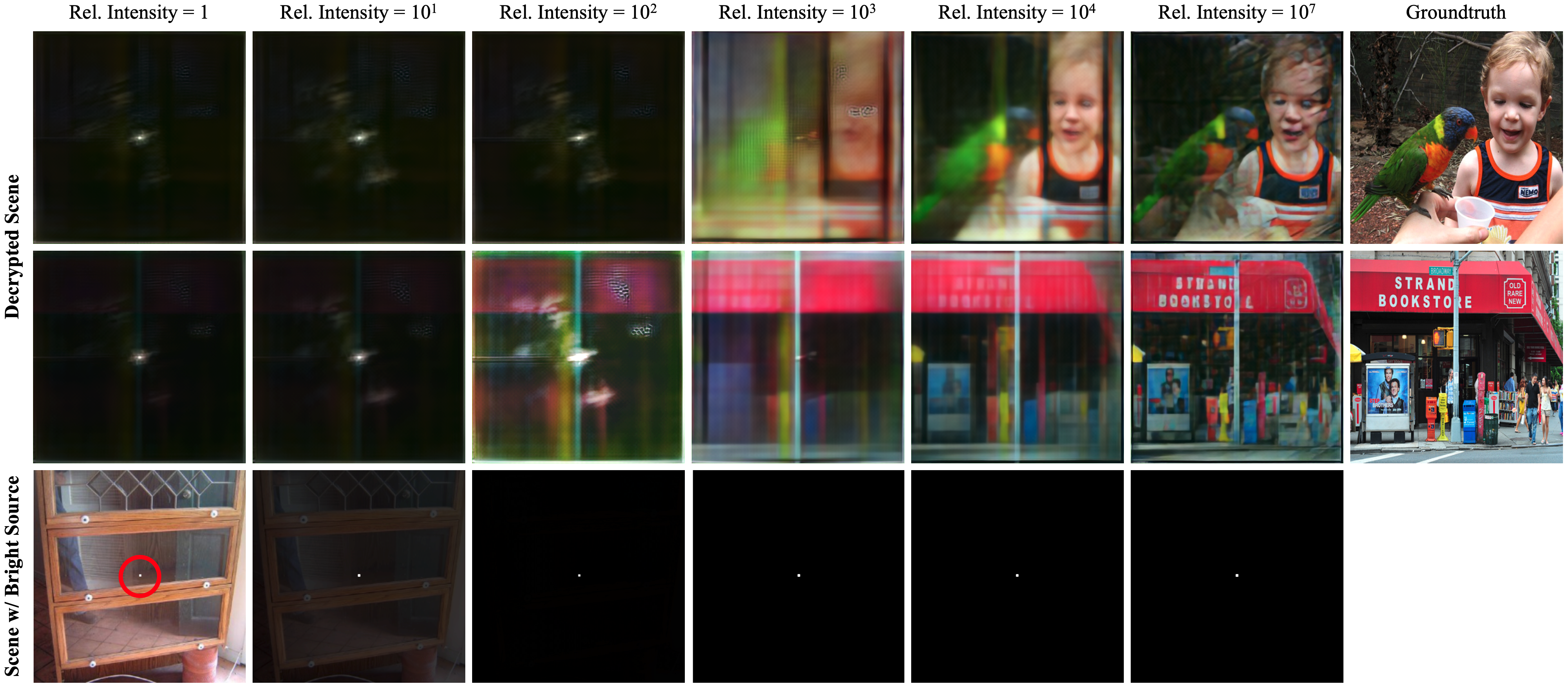}
    \caption{\textbf{ Uniform-Impulse - KPA.} (Top Rows) We show the decrypted scenes for varying relative intensities of the bright source. (Bottom Row) We show the scene with a point source (encircled) of varying relative intensity. \opencam offers privacy for a wide range of practical relative point source intensities.}
    \label{fig:white_wall}
\end{figure*}

In this attack, the attacker tries to jointly estimate the scaling mask $S$ and PSF $P$ and then uses them to decrypt the ciphertext. We assume the attacker has access to the approximate impulse response for a scene with a dominating bright source, and additional information in the form of a white-wall-measurement. We then solve the optimization,
\begin{equation}
    \{\hat{S},\hat{P}\} = \argmin_{S, P} \sum_{i=1}^{2}||Y_i - S\odot(P*X_i)||_F^2,
\end{equation}to estimate the keys. $Y_1$ is a USR, and $Y_2$ is the captured impulse response, while $X_1$ and $X_2$ are all-ones images and impulse images respectively. $Y_1$ can also be an average measurement corresponding to a large diverse set of natural scenes. Once $S$ and $P$ are estimated, the attacker uses the keyed-decryption framework of Section \ref{keyedec} to decrypt the ciphertexts/measurements. 
We show the results for different relative intensities of the point source with respect to the rest of the scene in fig. \ref{fig:white_wall}. For cases where the relative intensity of the bright source is less than $10^4$, it is not possible to accurately estimate the scaling mask and the PSF, and as a result, the quality of the decrypted scene is poor. For relative intensities beyond $10^4$, it is possible to reconstruct the scene to a reasonable extent. However, such large relative intensities of point sources imply dark-room-like environments which are less likely to occur in the attack model we have assumed i.e. the attacker doesn't have access to the physical location of the camera. Nevertheless, we acknowledge that this could be a limitation of \opencam and leave improving security for such scenarios as future work. 

%%%%%%%%%%%%%%%%%%%%%%%%%%%%%%%%%%%%%%%%%%%%%%%%%%%%%%%%%%%%%%%%%%%%%%%%%%
\subsection{Real Experimental Results}
%%%%%%%%%%%%%%%%%%%%%%%%%%%%%%%%%%%%%%%%%%%%%%%%%%%%%%%%%%%%%%%%%%%%%%%%%%

\begin{figure*}
    \centering
    \includegraphics[width=\linewidth]{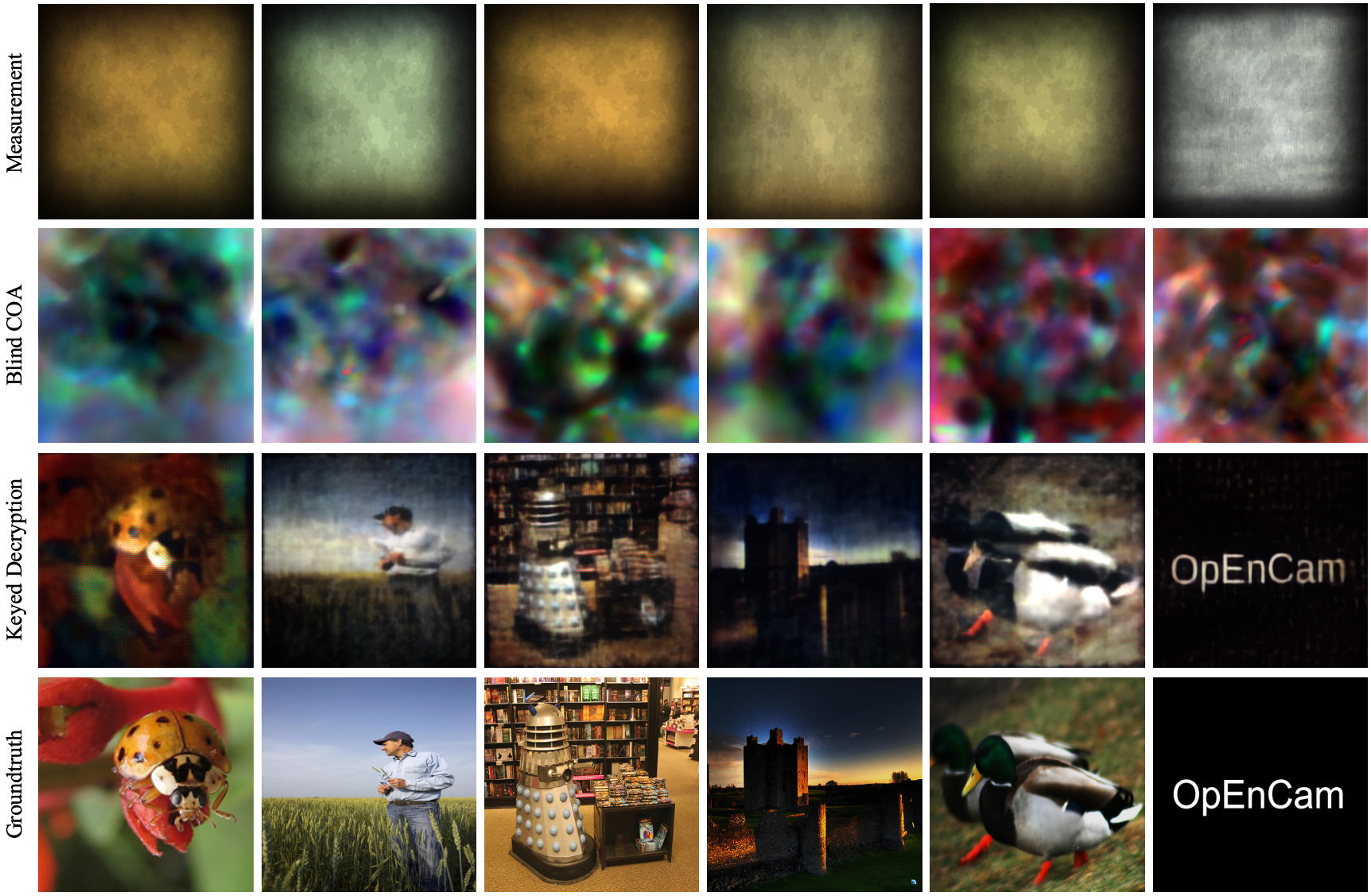}
    \caption{\textbf{Real Results from {OpEnCam} Prototype.} The top row shows {OpEnCam} measurements. The second row shows the reconstructions from a DPT-based blind COA. The third row shows keyed reconstructions. The bottom row shows the ground-truth scene.}
    % \vspace{-1.5em}
    \label{fig:real_results_supp}
\end{figure*}

% \begin{figure}[!ht]
%     \centering
%     \includegraphics[width=\linewidth]{figures/real_prot.png.jpg}
%     \caption{\textbf{OpEnCam Prototype.} We show our \opencam proof-of-concept prototype built from a hobby webcam and masks printed using office printers. The corresponding PSF and scaling mask are shown on the right.}
%     \label{fig:prot}
% \end{figure}
For the real-world experiments, we construct a working prototype {OpEnCam} using Basler Ace4024-29uc machine vision camera. To implement the multiplexing mask, we use an amplitude mask printed using a conventional office printer and place it over a Perlin contour phase mask. The scaling mask $S$ essentially implements a spatially varying exposure pattern for each sensor pixel in \opencam. Given that placing the scaling mask flush against the bare sensor requires breaking the sensor's protective covering, we implement its effect through exposure bracketing. That is, we quantize the scaling mask into 16 levels and then capture the measurement due to the multiplexing mask at 16 different relative exposure values given by the 16 levels of quantization. Finally, using the quantized scaling mask, we blend the measurement. We capture the experimental scaling mask ($S_{exp}$) by displaying a white image on a monitor and blending the measurements captured at 16 different exposure values without the multiplexing mask. We then capture the experimental PSF ($P_{exp}$) in the same way, but with the multiplexing mask, i.e., we capture the measurements for a point source at 16 different exposure values and blend them. The true PSF is then the normalized version of $P_{exp}$ i.e., $P_{true} = P_{exp} / (S_{exp}+\epsilon)$. Finally, we display images of natural scenes on the monitor, capture the corresponding exposure stack, and blend it to obtain the experimental measurements. For keyed decryption on these experimental data, we use the Tikhonov regularized reconstruction (Eq. (\ref{tikhonov})) to obtain the intermediate scene estimates. Since these estimates are generally noisy, we then use a denoising U-Net to refine them. For blind transformer-based attack, we use the DPT trained in Section \ref{coa_expt} on the experimental measurements. We show the visual results in fig. \ref{fig:real_results_supp}. These results demonstrate the efficacy of our approach -- recovery of the original scene from an {OpEnCam} measurement is possible only if one has access to the camera's encryption key.

%%%%%%%%%%%%%%%%%%%%%%%%%%%%%%%%%%%%%%%%%%%%%%%%%%%%%%%%%%%%%%%%%%%%%%%%%%
\section{Conclusion and Discussion}
%%%%%%%%%%%%%%%%%%%%%%%%%%%%%%%%%%%%%%%%%%%%%%%%%%%%%%%%%%%%%%%%%%%%%%%%%%

We propose a novel design for lensless cameras called \opencam to perform optical encryption. At the core of our method is the double-mask imaging model which implements shift-variant forward encryption process. Moreover, using ideas derived from signal processing, we design the mask patterns of our system to not only allow high-quality lensless imaging but also be robust against various forms of attacks. Through extensive experiments, we showed that our proposed \opencam design is robust against powerful learning-based ciphertext-only attacks and three special cases of known plaintext attacks. Finally, we build an \opencam prototype that allows us to validate our claims through promising preliminary results. Using a photolithography-based phase mask printing process\cite{boominathan2020phlatcam}, our prototype's performance can significantly improve, and we plan to implement this in the future.

Although preliminary results from \opencam show promising results, it should be noted that, like existing optical encryption systems, \opencam also has its limitations. Accessing the hardware or allowing extensive control of the scene has the potential to reveal the key. However, as pointed out through our experiments, the \opencam design is still an improvement over the existing optical encryption system and is a step towards more secure imaging.

In future, it would be interesting to look into the data-driven design of \opencam systems and using multiple-masked lensless cameras for other applications.
%%%%%%%%% REFERENCES

\bibliographystyle{IEEEtran}
\bibliography{refs}

\vfill

\end{document}